\documentclass[a4paper,12pt]{article}
\usepackage{amsmath,amssymb,bm}
\usepackage{graphicx}
\usepackage{enumerate}
\usepackage{mathrsfs}

\title{An Axiom for Concavifiable Preferences in View of Alt's Theory}
\author{Yuhki Hosoya\thanks{E-mail: hosoya(at)tamacc.chuo-u.ac.jp}\\ Faculty of Economics, Chuo University\thanks{742-1 Higashinakano, Hachioji-shi, Tokyo, 192-0393 Japan.}}
\date{\today}

\begin{document}
\maketitle

\begin{abstract}
We present a necessary and sufficient condition for Alt's system to be represented by a continuous utility function. Moreover, we present a necessary and sufficient condition for this utility function to be concave. The latter condition can be seen as an extension of Gossen's first law, and thus has an economic interpretation. Together with the above results, we provide a necessary and sufficient condition for Alt's utility to be continuously differentiable.

\vspace{12pt}

\vspace{12pt}

\end{abstract}

\section{Introduction}
In economics, there are a number of conditions that can usually be assumed for utility functions, such as continuity, quasi-concavity, and strict quasi-concavity. These conditions are accepted because the axioms of the preference relation that can be represented by a utility function satisfying such conditions are easy to understand and have a natural interpretation in economics. Differentiability, on the other hand, is unlike these. The conditions for a preference relation that can be represented by a differentiable utility function are neither easy to understand nor naturally interpretable in terms of economics. Often, however, the differentiability of utility functions is accepted simply because it is overwhelmingly convenient.

In this paper, we treat the concavity of utility functions. Concavity is not usually considered a good assumption for utility functions. There are several reasons for this, but the most important one is the difficulty of interpreting concavity in economic terms. A characterization of such a preference relation that can be represented by a concave utility function can be found in Kannai (1977). Three axioms are presented, corresponding to the condition of concavity without differentiation, with first-order derivatives, and with second-order derivatives. However, all of these axioms are too complicated for their meaning to be interpreted in economic terms. On the contrary, it is difficult to determine, even mathematically, whether or not these conditions are satisfied for a given preference relation. Because of this problem, the assumption that the utility function is concave is frowned upon, at least in theoretical research.

The advantages of assuming the concavity for the utility function are, however, too great to ignore. First, if the utility function is concave, then Lagrange's multiplier rule can be applied to the derivation of the demand function using subdifferential analysis.\footnote{See, for example, Rockafeller (1996) or Ioffe and Tikhomirov (1979) for detailed arguments.} This advantage is not negligible in real calculations. Second, there are some concepts that cannot be defined without the existence of a concave utility function. For example, Debreu's (1976) least concave utility is one of these concepts. Kannai (1980) connected this concept with ALEP substitution and complementarity in equilibrium theory. Epstein and Zhang (1999) showed that the order among concave utility representations that was introduced by Debreu (1976) is related to the concept of uncertainty aversion. Thus, there are many applications of the least concave utility function. Third, there are several properties that emerge from the presence of concave utility functions as the input to a Samuelson-Bergson type social welfare function, and concavity has a non-negligible effect on welfare analysis in applied research.

Let us elaborate a little on the last point. If one extends the preference relation from the usual space to the space of simple lotteries, then some results are known about the concave representability of the Neumann-Morgenstern utility function. However, the use of the Neumann-Morgenstern utility function in welfare analysis has long been criticized, by Luce and Raiffa (1957) and others, as being interpretatively unnatural. Furthermore, because the basic model of general equilibrium theory is not a stochastic model, the use of the Neumann-Morgenstern utility function is not desirable.

In this paper, we revisit the existence theorem of concave utility function from Alt's representation theory (Alt, 1936). To the best of our knowledge, this is the oldest theory to have discussed the possibility of utility representation of preference relations. Although this theory is also a kind of cardinal utility theory, it has an interpretation that is more suitable for welfare analysis than that of Neumann-Morgenstern utility. Furthermore, Alt's theory is not especially incompatible with general equilibrium theory. In these respects, it is worth deriving the necessary and sufficient conditions for Alt's utility function to be concave.

Note that the fact that Alt's utility function is concave is more important than just the fact that the preference is represented by some concave utility function. Because Alt's utility function has an interpretation that is suitable for welfare analysis, it can be treated as an input to the Samuelson-Bergson type social welfare function naturally. Recall that almost all typical Samuelson-Bergson type social welfare functions, including the Bentham, Rawls, and Nash types, are quasi-concave. Now, consider a quasi-concave social welfare function. Suppose that there are two individuals with the same utility function. If this utility function is concave, then bringing the states of these two individuals closer together does not decrease welfare. If the utility function is strictly concave, then bringing the states of these two individuals closer together will increase welfare in many cases. Thus, if the utility function is concave, the ethical principle that ``like people should be in like states'' is justified. In contrast, if the utility function is not concave, then it may be possible to improve welfare by putting similar people in different situations. Thus, when discussing welfare analysis, whether the utility function is concave or not has a decisive influence on the ethical principle. From this viewpoint, we think that the concavity of Alt's utility function is important for welfare analysis.

In this paper, we first derive a necessary and sufficient condition for the existence of a continuous utility function that represents Alt's system (Theorem 1). Although Alt essentially showed such a result, the axioms he used were too numerous and not independent, and were difficult to interpret economically. Similar problems can be found in more recent studies, such as Kranz et al. (1971). Seidl and Schmidt (1997) elaborated on this result, but again the number of conditions is too large and not independent. In contrast, Shapley (1975) derived an existence theorem for Alt's utility function with only three simple axioms. However, the space treated by Shapley is limited to a subset of the one-dimensional space. Our first result can be seen as an extension of Shapley's result to Alt's system on a Hausdorff, separable, and path-connected topological space.

We then derive a necessary and sufficient condition for Alt's utility function to be concave (Theorem 2). Because this condition can be seen as a generalization of Gossen's first law, we name this condition ``generalized Gossen's first law''. Compared with Kannai's axiom, this generalized Gossen's first law is much easier to interpret economically.

Finally, because Gossen's first law is usually described as a feature of partial derivatives of the utility function, we want to obtain the conditions for Alt's utility function to be differentiable. Hence, we present a necessary and sufficient condition for Alt's utility function to be continuously differentiable and nondegenerate (Theorem 3). This result is also useful for analysis of ALEP substitution and complementarity.

Note that generalized Gossen's first law is stronger than Kannai's axiom as a condition for the existence of concave utility functions. Our condition is a necessary and sufficient condition for Alt's utility function to be concave, whereas Kannai's axiom only guarantees the existence of a concave utility function to represent a given preference relation. We believe this axiom is still useful because it has a much more natural economic interpretation than Kannai's axiom. This is, in our opinion, one of the most important requirements when discussing whether or not to assume such a property for the utility function.

In subsection 2.1, we introduce the notion of Alt's system, and discuss its interpretation. In subsections 2.2 and 2.3, we present three axioms for Alt's system, and state that they are equivalent to the existence of the corresponding utility function. In subsections 2.4 and 2.5, we treat generalized Gossen's first law, and show that this axiom is equivalent to the concavity of Alt's utility function. In subsection 2.6, we examine the differentiability of Alt's utility. In section 3, we discuss the relationship between this work and several related studies. Because the proofs of Theorems 1 and 3 are somewhat lengthy, they are given in the appendix.

\section{Results}
\subsection{Preliminaries: Alt's System}
Let $X$ be a nonempty set. A binary relation $\ge$ on $X^2$ is called an \textbf{Alt's system} on $X$. A function $u:X\to \mathbb{R}$ is said to represent Alt's system $\ge$ (or to be a utility function of this system) if and only if the following requirement holds: for all $(x,y,z,w)\in X^4$,
\begin{equation}
u(x)-u(y)\ge u(z)-u(w)\Leftrightarrow (x,y,z,w)\in \ge.\label{REI}
\end{equation}
This relatiohship (\ref{REI}) says that the function $(x,y)\mapsto u(x)-u(y)$ represents $\ge$.

For an Alt's system $\ge$, we define
\begin{equation}\label{ORDER}
\succsim=\{(x,y)\in X^2|(x,y,y,y)\in \ge\}.
\end{equation}
If (\ref{REI}) holds for some function $u$, then
\begin{equation}\label{REU}
u(x)\ge u(y)\Leftrightarrow (x,y)\in \succsim.
\end{equation}
This relationship (\ref{REU}) says that $u$ is a utility function that represents $\succsim$.

The interpretation of Alt's system is as follows. $(x,y,z,w)\in \ge$ means that the \textbf{strength of the improvement} from $y$ to $x$ is not weaker than that from $w$ to $z$. If $u$ represents this system, then $u(x)-u(y)$ measures this strength, and thus $u$ is a sort of cardinal utility function. Clearly, if $u$ represents this system, then $(x,y)\in \succsim$ means that $x$ is preferred to $y$.

As usual, we write $x\succsim y$ instead of $(x,y)\in \succsim$. Moreover, we write $[x,y]\ge [z,w]$ instead of $(x,y,z,w)\in \ge$.\footnote{Alt originally used the arrow notation $[y\to x]\ge [w\to z]$. On the other hand, recent related papers (e.g. Wakker (1988), Miyake (2016), Gerasimou (2021)) usually use the following simple notation $(x,y)\ge (z,w)$. However, the notation $(x,y)$ is not desirable in this paper because we later treat elements of $\mathbb{R}^n$, and thus this abstract pair $(x,y)$ may be mistaken for an element of $\mathbb{R}^2$. Hence, we use $[x,y]$ instead of $(x,y)$.} These notations allow us to understand what is occurring. We write $x\succ y$ if $x\succsim y$ and $y\not\succsim x$ and $x\sim y$ if $x\succsim y$ and $y\succsim x$. Additionally, we write $[x,y]=[z,w]$ if $[x,y]\ge [z,w]$ and $[z,w]\ge [x,y]$, and $[x,y]>[z,w]$ if $[x,y]\ge [z,w]$ and $[z,w]\not\ge [x,y]$.

\subsection{Axioms}
We now present several axioms on Alt's system. The first one is necessary for the system to match its interpretation appropriately.

\vspace{12pt}
\noindent
\textbf{Definition 1}. An Alt's system $\ge$ is said to satisfy \textbf{consistency} if and only if, for every $x,y,z\in X$,
\[x\succsim y\Leftrightarrow [x,z]\ge [y,z].\]

\vspace{12pt}
Note that, under this consistency axiom, if there exists $z\in X$ such that $[x,z]\ge [y,z]$, then $x\succsim y$. Conversely, if $x\succsim y$, then \textbf{for every} $z$, $[x,z]\ge [y,z]$. This gap (between `there exists' and `for every') may be a source of confusion.

Under consistency, we can prove that $\succsim$ is a weak order on $X$. First, choose any $x,y\in X$. If $x\not\succsim y$, then $[x,y]\not\ge [y,y]$, and thus $[y,y]\ge [x,y]$. By consistency, we have that $y\succsim x$, which implies that $\succsim$ is complete. Second, suppose that $x\succsim y$ and $y\succsim z$. Then,
\[[x,y]\ge [y,y],\ [y,z]\ge [z,z].\]
Because of consistency, $[x,z]\ge [y,z]$, and thus $[x,z]\ge [z,z]$ by transitivity of $\ge$, which implies that $x\succsim z$, and thus $\succsim$ is transitive. Therefore, if $\ge$ satisfies consistency, then $\succsim$ is a weak order.

If there is a representation $u$ of $\ge$, then consistency implies that
\[u(x)\ge u(y)\Leftrightarrow u(x)-u(z)\ge u(y)-u(z),\]
and thus, consistency is trivially satisfied.

The next axiom relates to the property of $=$.

\vspace{12pt}
\noindent
\textbf{Definition 2}. An Alt's system $\ge$ is said to satisfy the \textbf{crossover axiom} if and only if, for every $x,y,z,w\in X$,\footnote{The name `the crossover axiom' was used in Miyake (2016).}
\[[x,y]=[z,w]\Leftrightarrow [x,z]=[y,w].\]

\vspace{12pt}
The relation $=$ is symmetric, and therefore the crossover axiom also means that
\[[x,y]=[z,w]\Leftrightarrow [w,y]=[z,x].\]
If there is a representation $u$ of $\ge$, then the crossover axiom implies that
\[u(x)-u(y)=u(z)-u(w)\Leftrightarrow u(x)-u(z)=u(y)-u(w),\]
and thus, this axiom is also trivially satisfied.

We note two facts. First, for every $x,y\in X$,
\[[x,y]=[x,y].\]
If the crossover axiom holds, then this implies that
\[[x,x]=[y,y].\]
That is, ``the strength of the improvement under the unchanged situation is the same at every point.'' This is quite natural. Second, under consistency and the crossover axiom,
\[x\sim y\Leftrightarrow [x,z]=[y,z]\Leftrightarrow [x,y]=[z,z]\]
for every $x,y,z\in X$. That is, ``$x$ is indifferent to $y$ if and only if the strength of the improvement from $y$ to $x$ is the same as that in the unchanged situation.'' This is also natural. 

To treat the third axiom, we must consider $X$ as a topological space.

\vspace{12pt}
\noindent
\textbf{Definition 3}. Suppose that $X$ is a topological space. Then, an Alt's system $\ge$ is said to satisfy \textbf{continuity} if and only if $\ge$ is closed in $X^4$.

\vspace{12pt}
Note that, if $\ge$ satisfies consistency and continuity, then $\succsim$ is also closed in $X^2$ by (\ref{ORDER}).

If there is a continuous representation $u$ of $\ge$, then clearly $\ge$ is closed, because
\[\ge=\{(x,y,z,w)\in X^4|u(x)+u(w)-u(y)-u(z)\ge 0\}.\]

\subsection{Result for the Existence of Alt's Cardinal Utility}
The next theorem is our first result.

\vspace{12pt}
\noindent
\textbf{Theorem 1}. Suppose that $X$ is a Hausdorff, separable, and path-connected topological space, and $\ge$ is an Alt's system on $X$.\footnote{A topological space $X$ is said to be \textbf{path-connected} if and only if for every $x,y\in X$, there exists a continuous function $f:[0,1]\to X$ such that $f(0)=x,\ f(1)=y$. For example, every convex set of some topological vector space is path-connected. It is well known that every path-connected space is connected.

Note that, we do not know whether Theorem 1 still holds on a connected topological space that is not path-connected. See section 3 and footnote 16 for more detailed arguments.} Then, there exists a continuous representation $u:X\to \mathbb{R}$ of $\ge$ if and only if this system satisfies consistency, the crossover axiom, and continuity. If so, such a representation is unique up to a positive affine transform: that is, if $u_1,u_2$ are continuous representations of $\ge$, then there exist $a>0$ and $b\in \mathbb{R}$ such that
\[u_2(x)=au_1(x)+b\]
for every $x\in X$.

\vspace{12pt}
The proof of this theorem is in the appendix.

\subsection{Generalized Gossen's First Law}
Our main purpose is to present a necessary and sufficient condition for $u$ to be concave. Hence, we introduce the following axiom.

\vspace{12pt}
\noindent
\textbf{Definition 4}. Suppose that $X$ is a separable and convex subset of a Hausdorff topological vector space. Then, an Alt's system $\ge$ on $X$ is said to satisfy \textbf{generalized Gossen's first law} if and only if, for every $x,y\in X$,
\[[z,x]\ge [y,z]\]
for $z=\frac{1}{2}(x+y)$. If, in addition,
\[[z,x]>[y,z]\]
whenever $x\neq y$, then $\ge$ is said to satisfy \textbf{generalized strong Gossen's first law}.

\vspace{12pt}
We should interpret the meaning of generalized Gossen's first law. Suppose that $X$ is a convex subset of $\mathbb{R}^n$, and $x\in X$ and $y=(x_1+2a,x_2,...,x_n)\in X$ for $a>0$. Then, $z=(x_1+a,x_2,...,x_n)\in X$. Suppose that generalized Gossen's first law is satisfied. Then, $[z,x]\ge [y,z]$. If $u$ is a continuously differentiable representation of this system, then by the mean value theorem,
\[u(z)-u(x)=a\frac{\partial u}{\partial x_1}(z-\theta ae_1),\]
\[u(y)-u(z)=a\frac{\partial u}{\partial x_1}(z+\eta ae_1)\]
for some $\theta,\eta\in [0,1]$, where $e_1=(1,0,...,0)$. Thus, roughly speaking, $[z,x]\ge [y,z]$ implies that the mapping 
\[b\mapsto \frac{\partial u}{\partial x_1}(x_1+b,x_2,...,x_n)\]
is nonincreasing around $a$. Because $x$ and $a$ are arbitrary, we conclude that under generalized Gossen's first law, the marginal utility of $x_1$ is nonincreasing in $x_1$. This explains why this axiom is named as such.

In the definition of generalized Gossen's first law, $z=x+(z-x)$ and $y=z+(z-x)$. Therefore, this axiom can be transformed as follows: for every $v\neq 0$ such that $x+2v\in X$,
\[[x+v,x]\ge [x+2v,x+v].\]
If $u$ represents this system, then
\[u(x+v)-u(x)\ge u(x+2v)-u(x+v),\]
and thus, under this axiom, the marginal improvement of the utility for a change of consumption in the $v$ direction decreases as the length of the change increases.

\subsection{Result for the Concavity of Alt's Cardinal Utility}
The following is our second main result.

\vspace{12pt}
\noindent
\textbf{Theorem 2}. Suppose that $X$ is a separable and convex subset of a Hausdorff topological vector space, $\ge$ is an Alt's system on $X$, and $u:X\to \mathbb{R}$ is a continuous function that represents this system. Then, $u$ is concave (resp. strictly concave) if and only if $\ge$ satisfies generalized Gossen's first law (resp. generalized strong Gossen's first law).

\vspace{12pt}
Note that, because $X$ is convex, it is path-connected, and thus all assumptions in Theorem 1 are satisfied.

\vspace{12pt}
\noindent
\textbf{Proof}. First, suppose that $u$ is concave. Choose any $x,y\in X$, and define $z=\frac{1}{2}(x+y)$. Then,
\begin{equation}
\frac{1}{2}u(z)+\frac{1}{2}u(z)=u(z)\ge \frac{1}{2}u(x)+\frac{1}{2}u(y),\label{GFL}
\end{equation}
and thus
\[u(z)-u(x)\ge u(y)-u(z),\]
as desired. If, in addition, $u$ is strictly concave and $y\neq x$, then the inequality in (\ref{GFL}) is strengthened, and thus
\[u(z)-u(x)>u(y)-u(z),\]
as desired.

Conversely, suppose that $\ge$ satisfies generalized Gossen's first law. We want to show that
\begin{equation}
u((1-t)x+ty)\ge (1-t)u(x)+tu(y)\label{concave}
\end{equation}
for every $x,y\in X$ and $t\in [0,1]$. Choose any $x,y\in X$, and define $z=\frac{1}{2}(x+y)$. By generalized Gossen's first law,
\[[z,x]\ge [y,z],\]
and thus,
\[u(z)-u(x)\ge u(y)-u(z),\]
which implies that
\[u(z)\ge \frac{1}{2}u(x)+\frac{1}{2}u(y).\]
Therefore, (\ref{concave}) holds if $t=\frac{1}{2}$. By mathematical induction, we can easily show that (\ref{concave}) holds if $t$ is a dyadic rational $\frac{m}{2^{\ell}}$, where $\ell\ge 1$ and $1\le m\le 2^{\ell}-1$. Because the set of all dyadic rationals is dense in $[0,1]$, by the continuity of $u$, (\ref{concave}) holds for all $t\in [0,1]$.

Suppose that $\ge$ satisfies generalized strong Gossen's first law. Choose any $x,y\in X$ such that $x\neq y$. Define $z=\frac{1}{2}(x+y)$. Then,
\[[z,x]>[y,z],\]
and thus,
\[u(z)-u(x)>u(y)-u(z),\]
which implies that
\[u(z)>\frac{1}{2}u(x)+\frac{1}{2}u(y).\]
Therefore,
\begin{equation}
u((1-t)x+ty)>(1-t)u(x)+tu(y)\label{sconcave}
\end{equation}
if $t=\frac{1}{2}$. By mathematical induction, we can easily show that (\ref{sconcave}) holds if $t$ is a dyadic rational $\frac{m}{2^{\ell}}$, where $\ell\ge 1$ and $1\le m\le 2^{\ell}-1$. Now, choose any $t\in ]0,1[$ such that $t$ is not a dyadic rational. Then, if $\ell$ is sufficiently large, there exists $m\in \{1,...,2^{\ell}-2\}$ such that $t_0=\frac{m}{2^{\ell}}<t<\frac{m+1}{2^{\ell}}=t_1$. In this case,
\[t=\frac{t_1-t}{t_1-t_0}t_0+\frac{t-t_0}{t_1-t_0}t_1,\]
and thus, by the continuity of $u$,
\begin{align*}
u((1-t)x+ty)\ge&~\frac{t_1-t}{t_1-t_0}u((1-t_0)x+t_0y)+\frac{t-t_0}{t_1-t_0}u((1-t_1)x+t_1y)\\
>&~\frac{t_1-t}{t_1-t_0}[(1-t_0)u(x)+t_0u(y)]\\
&~+\frac{t-t_0}{t_1-t_0}[(1-t_1)u(x)+t_1u(y)]\\
=&~(1-t)u(x)+tu(y),
\end{align*}
as desired. This completes the proof. $\blacksquare$

\subsection{Result for the Differentiability of Alt's Cardinal Utility}
Usually, Gossen's laws are stated in the language of differential calculus. Hence, we provide a sufficient condition for Alt's utility to be differentiable.

First, let $x,y\in\mathbb{R}^n$. We write $x\ge y$ if and only if $x_i\ge y_i$ for all $i\in \{1,...,n\}$, and $x\gg y$ if and only if $x_i>y_i$ for all $i\in \{1,...,n\}$. Define $\mathbb{R}^n_+=\{x\in \mathbb{R}^n|x\ge 0\}$ and $\mathbb{R}^n_{++}=\{x\in \mathbb{R}^n|x\gg 0\}$. We introduce three axioms.

\vspace{12pt}
\noindent
\textbf{Definition 5}. Let $X\subset \mathbb{R}^n$ and $\ge$ be an Alt's system on $X$. Then, this system is said to be \textbf{monotone} if $x\succ y$ for every $x,y\in X$ such that $x\gg y$.

\vspace{12pt}
\noindent
\textbf{Definition 6}. Let $X$ be either $\mathbb{R}^n_+$ or $\mathbb{R}^n_{++}$, and $\ge$ be an Alt's system on $X$. Then, this system is said to satisfy \textbf{Debreu's smoothness} if and only if the following set
\[I=\{(x,y)\in \mathbb{R}^{2n}_{++}|x\sim y\}\]
is a $2n-1$ dimensional $C^1$ manifold.

\vspace{12pt}
\noindent
\textbf{Definition 7}. Let $X$ be either $\mathbb{R}^n_+$ or $\mathbb{R}^n_{++}$, and $\ge$ be an Alt's system on $X$ that satisfies consistency, the crossover axiom, continuity, and monotonicity. Define $e=(1,1,...,1)\in X$. By Theorem 1 and the intermediate value theorem, for every $a,b>0$ with $a<b$, there uniquely exists $f(a,b)>0$ such that\footnote{The existence of such an $f$ can also be shown using Lemma 4 in the appendix.}
\[[f(a,b)e,(b-a)e]=[(b+a)e,f(a,b)e].\]
We say that this system satisfies \textbf{line smoothness} if and only if
\[\lim_{a\downarrow 0}\frac{b-f(a,b)}{a}=0\]
for all $b>0$.

\vspace{12pt}
\noindent
\textbf{Theorem 3}. Let $X$ be either $\mathbb{R}^n_+$ or $\mathbb{R}^n_{++}$, and $\ge$ be an Alt's system on $X$ that satisfies consistency, the crossover axiom, continuity, generalized Gossen's first law, and monotonicity. Suppose that $u$ is a continuous representation of $\ge$. Then, $u$ is continuously differentiable on $\mathbb{R}^n_{++}$ and $Du(x)\neq 0$ for every $x\in \mathbb{R}^n_{++}$ if and only if $\ge$ satisfies both Debreu's smoothness and line smoothness.

\vspace{12pt}
The proof of this theorem is in the appendix. Note that Debreu's smoothness does not imply line smoothness. For example, let $v(x)=\sqrt{x_1x_2}$ and 
\[g(c)=\begin{cases}
c-1 & \mbox{if }c\le 1,\\
\frac{c-1}{2} & \mbox{if }c>1.
\end{cases}\]
Define $u(x)=g(v(x))$, and suppose that $\ge$ is an Alt's system that corresponds to $u$. Then, this system satisfies Debreu's smoothness, because $I=\{(x,y)\in \mathbb{R}^{2n}_{++}|v(x)=v(y)\}$. However, $f(a,1)=1-\frac{1}{4}a$, and thus,
\[\lim_{a\downarrow 0}\frac{1-f(a,1)}{a}=\frac{1}{4}\neq 0,\]
which implies that this system violates line smoothness.

On the other hand, every function that is increasing and homogeneous of degree one satisfies line smoothness, although its indifference curve is kinked. Therefore, line smoothness does not imply Debreu's smoothness. That is, Debreu's smoothness is independent of line smoothness.

We should mention Alexandrov's theorem. This theorem states that for every convex function, its subdifferential mapping is `differentiable' in the sense of set-valued differentiation almost everywhere.\footnote{See Alexandrov (1939), section 6 of Howard (1998), or theorem 3.12.3 of Niculescu and Persson (2018) for detailed arguments.} Applying this theorem, we can obtain the following result: if $u$ is a concave and continuously differentiable utility function derived from Theorem 3, then it is twice differentiable almost everywhere. Thus, ALEP substitution and complementarity can be defined almost everywhere. 

Recall the definition of ALEP substitution and complementarity. This notion uses second cross-derivatives of the utility function. If $\frac{\partial^2u}{\partial x_i\partial x_j}<0$ (resp. $\frac{\partial^2 u}{\partial x_i\partial x_j}>0$), then commodity $i$ is a substitute (resp. complementary) good of commodity $j$. In other words, commodity $i$ is a substitute good of commodity $j$ if and only if an increase in consumption of commodity $i$ undermines the attractiveness of commodity $j$. This notion was argued by von Auspitz and Lieben (1889), Edgeworth (1897), and Pareto (1906), and the name `ALEP' comes from the first letters of these researchers' surnames.

Clearly, this concept only works if the difference in the value of $u$ implies the intensity of the improvement. Therefore, Alt's utility theory is very useful for this concept. Samuelson (1974) discussed in detail the definition of ALEP complementarity using cardinal utility, and argued the relationship between this notion and Alt's utility. This discussion is also explained in Kawamata (2010). The fact mentioned in the paragraph above means that by using our Theorem 3, we can discuss the substitution-complementarity relationship in the sense of ALEP to Alt's utility. In this sense, Theorem 3 can be considered as adding more value to Alt's utility.

\section{Remarks}
To the best of our knowledge, Alt's system was first argued by Alt (1936), who proved that under several axioms (including our consistency and continuity, and moreover, concatenation axiom discussed later), there exists a continuous representation. He wrote his paper in German, and it was translated into English in 1971 (Chipman et al. (1971)). Actually, his axioms include several `facts' that can be shown from our three axioms. For example, our Lemma 5 is one of Alt's axioms. Similar axioms can be found in Theorem 4.2 of Kranz et al. (1971), where our Lemmas 2 and 5 are treated as axioms. Shapley (1975) used only our three axioms and derived a continuous representation.\footnote{Shapley defined Alt's system as a pair $(\succsim,\ge)$ of weak orders on $X$ and $X^2$. However, using our definition of $\succsim$ (see (\ref{ORDER})), there is no difference between Shapley's setup and ours.} In his proof, however, $X$ is treated as a convex subset of the real line. Hence, our Theorem 1 is an extension of his result to a general path-connected topological space.

Theorem 5.3 of Wakker (1988) treats similar axioms to this paper and derives a similar result to our Theorem 1 on a connected topological space. His axioms are continuity and the following two axioms: 1) ({\it reversal condition}) If $[x,y]\ge [z,w]$, then $[w,y]\ge [z,x]$. 2) ({\it concatenation axiom}) $[x,y]\ge [x',y']$ and $[y,z]\ge [y',z']$ imply $[x,z]\ge [x',z']$. First, reversal condition is stronger than the crossover axiom, because the crossover axiom only states that reversal condition holds for not $\ge$ but only $=$. Second, under reversal condition and continuity, we can easily derive consistency. Therefore, our axioms are weaker than Wakker's axioms. Instead, we need not only connectedness, but also path-connectedness of the space, and thus we think that our Theorem 1 and Wakker's result is independent.\footnote{We guess that, under our three axioms, Lemma 7 is essentially needed to show that concatenation axiom holds. Later, we will see that, in proving Theorem 1, only Lemma 7 requires the path-connectedness. Therefore, we think that the reason why Wakker's result does not need path-connectedness is assuming concatenation axiom directly.}

Miyake (2016) considered an Alt's system on $\mathbb{R}^n_{++}$, and showed that, under several axioms, there exists a representation $\log u(x)$ such that $u$ is homogeneous of positive degree. Surprisingly, his result does not use the crossover axiom. Therefore, his result is independent of our Theorem 1.

Gerasimou (2021) showed that, under several axioms, our $\ge$ is represented by a general preference intensity function $g(x,y)$. His axioms are weaker than our axioms in Theorem 1. Instead, under our axioms, we can take $g(x,y)=u(x)-u(y)$, which is stronger than Gerasimou's result. Therefore, our Theorem 1 is independent of Gerasimou's result.

We should mention our assumption of path-connectedness. A topological space $X$ is said to be \textbf{connected} if and only if, for every pair of nonempty closed sets $A,B$ in $X$, either $A\cup B\neq X$ or $A\cap B\neq \emptyset$. Debreu (1954) showed that if $X$ is a Hausdorff, separable, and connected topological space, and a weak order $\succsim$ on $X$ is closed in $X^2$, then there exists a continuous function $u:X\to \mathbb{R}$ such that
\[u(x)\ge u(y)\Leftrightarrow x\succsim y.\]
Compared with this result, our Theorem 1 uses path-connectedness instead of connectedness. It is known that every path-connected topological space is connected. On the other hand, it is also known that the following set
\[X=\{(x^{-1},\sin x)|x\neq 0\}\cup \{(0,y)|-1\le y\le 1\}\]
is closed and connected, but not path-connected. Hence, path-connectedness is stronger than connectedness. In the appendix, path-connectedness is only used in the proof of Lemma 7, and in the remaining parts of this proof, only connectedness is used. Therefore, if Lemma 7 can be proved without the use of path-connectedness, we could extend Theorem 1 for a general connected space. However, at least we are not currently able to prove Lemma 7 without the aid of path-connectedness.

Debreu (1976) treated the notion of the least concave utility function. Suppose that $X$ is a separable and convex set of a Hausdorff topological vector space, and $\succsim$ is a weak order of $X$ that is closed in $X^2$. Let $\mathscr{U}$ be the set of all continuous and concave functions $u$ such that
\[u(x)\ge u(y)\Leftrightarrow x\succsim y.\]
If $u,v\in \mathscr{U}$, then we write $u\succeq v$ if and only if there exists a concave function $\varphi$ such that $u=\varphi\circ v$. Debreu showed that if $\mathscr{U}$ is nonempty, then there exists a least element in $\mathscr{U}$ with respect to this partial order $\succeq$. This least element is called the \textbf{least concave utility function}. He showed that the least concave utility function is unique up to a positive affine transform, and for fixed $x\in X$, if there exists a continuously differentiable element $u\in \mathscr{U}$ at $x$, then the least concave utility function is also continuously differentiable at $x$. The proof is very similar to Benveniste-Scheinkman's envelope theorem. In this paper, we characterized the existence of such a continuously differentiable $u\in \mathscr{U}$ in Theorem 3.

\appendix
\section{Proofs}
\subsection{Proof of a Lemma}
In this subsection, we introduce a property of an Alt's system $\ge$ on $X$. This system is said to satisfy \textbf{second consistency} if and only if for every $x,y,z\in X$,
\[x\succsim y\Leftrightarrow [z,y]\ge [z,x].\]
We show the following lemma.

\vspace{12pt}
\noindent
\textbf{Lemma 1}. Suppose that $X$ is a Hausdorff and connected topological space and $\ge$ is an Alt's system on $X$ that satisfies consistency, the crossover axiom, and continuity. Then, this system satisfies second consistency.

\vspace{12pt}
\noindent
\textbf{Proof}. We first show the following fact. Suppose that
\[[x_2,y]\ge [z,w]\ge [x_1,y].\]
Then, there exists $x_3\in X$ such that
\[[x_3,y]=[z,w].\]
To show this, let
\[U=\{x\in X|[x,y]\ge [z,w]\},\ D=\{x\in X|[z,w]\ge [x,y]\}.\]
Because $x_2\in U$ and $x_1\in D$, both $U$ and $D$ are nonempty. By continuity, both $U$ and $D$ are closed. Since $U\cup D=X$, by the connectedness of $X$, there exists $x_3\in U\cap D$. Then, clearly
\[[x_3,y]=[z,w].\]
Note that,
\[[x_2,y]\ge [x_3,y]\ge [x_1,y],\]
and thus, by consistency, $x_2\succsim x_3\succsim x_1$.

Second, suppose that $x\sim y$. Then, by consistency,
\[[x,z]=[y,z].\]
Therefore, by the crossover axiom,
\[[x,y]=[z,z],\]
and again by the crossover axiom,
\[[z,y]=[z,x].\]
Hence, second consistency holds for every $x,y,z\in X$ such that $x\sim y$.

Now, suppose that second consistency is violated. Then, by the above arguments, there exists $x,y,z\in X$ such that
\begin{equation}
y\succ x,\ [z,y]\ge [z,x].\label{VSC}
\end{equation}
If $y\succ z$, then by consistency, the crossover axiom, and (\ref{VSC}),
\[[x,x]=[y,y]>[z,y]\ge [z,x].\]
Therefore, there exists $w\in X$ such that
\[[w,x]=[z,y],\ x\succsim w\succsim z.\]
By the crossover axiom,
\[[w,z]=[x,y].\]
By consistency and the crossover axiom,
\[[z,z]=[y,y]>[x,y]=[w,z]\ge [z,z],\]
which is a contradiction.

Therefore, $z\succsim y$. By the transitivity of $\succsim$, we have that $z\succ x$. Thus, by (\ref{VSC}), consistency, and the crossover axiom,
\[[z,y]\ge [z,x]>[x,x]=[y,y].\]
Thus, there exists $w\in X$ such that
\[[w,y]=[z,x],\ z\succsim w\succsim y.\]
By the crossover axiom,
\[[w,z]=[y,x].\]
By the crossover axiom and consistency,
\[[z,z]\ge [w,z]=[y,x]>[x,x]=[z,z],\]
which is a contradiction. This completes the proof. $\blacksquare$

\subsection{Proof of Theorem 1}
Suppose that $X$ is a Hausdorff, separable, and path-connected topological space and $\ge$ is an Alt's system on $X$. If there exists a continuous function $u:X\to \mathbb{R}$ that represents $\ge$, then clearly $\ge$ satisfies consistency, the crossover axiom, and continuity. Therefore, it suffices to show the opposite direction, in other words, that if $\ge$ satisfies consistency, the crossover axiom, and continuity, then there exists a continuous function $u:X\to \mathbb{R}$ that represents $\ge$.

First, we introduce several lemmas.

\vspace{12pt}
\noindent
\textbf{Lemma 2}. If
\[[x_2,y]\ge [z,w]\ge [x_1,y],\]
then there exists $x_3\in X$ such that
\begin{equation}
[x_3,y]=[z,w].\label{L1}
\end{equation}
Moreover, $x_2\succsim x_3\succsim x_1$, and if $x_4$ also satisfies (\ref{L1}), then $x_3\sim x_4$.

\vspace{12pt}
\noindent
\textbf{Proof}. We have already proved this result in the proof of Lemma 1, except for the last claim. Hence, suppose that
\[[x_4,y]=[z,w].\]
By the transitivity of $=$,
\[[x_3,y]=[x_4,y]\]
and by consistency,
\[x_3\sim x_4,\]
as desired. This completes the proof. $\blacksquare$

\vspace{12pt}
\noindent
\textbf{Lemma 3}. Suppose that 
\[[y,x_1]\ge [z,w]\ge [y,x_2].\]
Then, there exists $x_3\in X$ such that
\begin{equation}
[y,x_3]=[z,w].\label{L2}
\end{equation}
Moreover, $x_2\succsim x_3\succsim x_1$, and if $x_4$ also satisfies (\ref{L2}), then $x_3\sim x_4$.

\vspace{12pt}
\noindent
\textbf{Proof}. By using second consistency instead of consistency, we can show this result using almost the same arguments as in the proof of Lemma 2. Hence, we omit the proof. $\blacksquare$

\vspace{12pt}
\noindent
\textbf{Lemma 4}. Suppose that $z\succ x$. Then, there exists $y\in X$ such that
\begin{equation}
[y,x]=[z,y].\label{L3}
\end{equation}
Moreover, $z\succ y\succ x$, and if $y'$ also satisfies (\ref{L3}), then $y\sim y'$.

\vspace{12pt}
\noindent
\textbf{Proof}. Define
\[U=\{w\in X|[w,x]\ge [z,w]\},\ D=\{w\in X|[z,w]\ge [w,x]\}.\]
Then, both $U$ and $D$ are closed. Because $[z,x]>[x,x]=[z,z]$, $z\in U$ and $x\in D$, and thus both $U$ and $D$ are nonempty. Because $U\cup D=X$, by the connectedness of $X$, there exists $y\in U\cap D$. Clearly,
\[[y,x]=[z,y].\]
If $y\succsim z$, then $y\succ x$, and thus
\[[y,x]\ge [z,x]>[z,y]=[y,x],\]
which is absurd. Therefore, $z\succ y$. Symmetrically, we can show that $y\succ x$.

Choose any $y'\in X$. If $y\succ y'$, then by consistency and second consistency,
\[[z,y']>[z,y]=[y,x]>[y',x].\]
If $y'\succ y$, then again by consistency and second consistency,
\[[y',x]>[y,x]=[z,y]>[z,y'].\]
Therefore, if $[z,y']=[y',x]$, then we must have $y\sim y'$. This completes the proof. $\blacksquare$

\vspace{12pt}
\noindent
\textbf{Lemma 5}.\footnote{Note that in the proof of Lemmas 1-4, we only used the connectedness of $X$; the separability and path-connectedness of $X$ were not used. However, in the proof of Lemma 5, we must use Debreu's representation theorem, and so $X$ must be separable.} Suppose that $x\succ y$. If $z\succsim x$, then there exists a finite sequence $a_0,a_1,...,a_k\in X$ such that\footnote{If there is a representation $u:X\to \mathbb{R}$, then this statement says that $(k-1)(u(x)-u(y))\le u(z)-u(x)<k(u(x)-u(y))$ for some $k\in \mathbb{N}$.}
\[a_0=y,\ a_1=x,\ [a_{i+1},a_i]=[a_{i+2},a_{i+1}]\mbox{ for all }i,\]
\[z\succsim a_k,\ [a_1,a_0]>[z,a_k].\]
Similarly, if $y\succsim w$, then there exists a finite sequence $a_1,a_0,...,a_{-k}$ such that
\[a_0=y,\ a_1=x,\ [a_{i+1},a_i]=[a_{i+2},a_{i+1}]\mbox{ for all }i,\]
\[a_{-k}\succsim w,\ [a_1,a_0]>[a_{-k},w].\]

\vspace{12pt}
\noindent
\textbf{Proof}. We treat only the claim on $z$, because the claim on $w$ can be proved symmetrically. Suppose that such a finite sequence does not exist. Define $a_0=y$ and $a_1=x$. By the negation of the claim of this lemma for $k=1$, we must have 
\[[z,a_1]\ge [a_1,a_0]>[a_1,a_1].\]
By Lemma 2, there exists $a_2\in X$ such that
\[[a_2,a_1]=[a_1,a_0],\ z\succsim a_2\succ a_1.\]
We assume that $a_0,...,a_k$ is already defined, and
\[[a_1,a_0]=[a_{i+1},a_i]\]
for every $i\in \{1,...,k-1\}$. By the negation of the claim of this lemma, we must have
\[[z,a_k]\ge [a_1,a_0]>[a_k,a_k].\]
Therefore, by Lemma 2, there exists $a_{k+1}\in X$ such that
\[[a_{k+1},a_k]=[a_1,a_0],\ z\succsim a_{k+1}\succ a_k.\]
Hence, by mathematical induction, we obtain an infinite sequence $(a_k)$ such that $z\succsim a_k$ and
\[[a_1,a_0]=[a_{k+1},a_k]\]
for every $k\in\mathbb{N}$.

We will show that
\begin{equation}
[a_{i+k},a_i]=[a_{j+k},a_j]\label{EQUIV}
\end{equation}
for every $i,j,k$. If $k=1$, then it has already been proved. By the crossover axiom,
\[[a_j,a_i]=[a_{j+1},a_{i+1}].\]
By the transitivity of $=$, for every $k$,
\[[a_j,a_i]=[a_{j+k},a_{i+k}].\]
Again by the crossover axiom,
\[[a_{i+k},a_i]=[a_{j+k},a_j],\]
and thus (\ref{EQUIV}) holds.

Now, by Debreu's representation theorem,\footnote{This theorem states that if $X$ is a Hausdorff, separable, and connected topological space and $\succsim$ is a weak order on $X$ that is closed in $X^2$, then there exists a continuous function $v:X\to \mathbb{R}$ such that 
\[v(x)\ge v(y)\Leftrightarrow x\succsim y\]
for every $x,y\in X$. For the proof of this theorem, see, for example, Debreu (1954) or Bridges and Mehta (1995).} there exists a continuous function $v:X\to \mathbb{R}$ such that
\[v(x')\ge v(y')\Leftrightarrow x'\succsim y'\]
for every $x',y'\in X$. Let $c^*=\sup_kv(a_k)$. Then, $v(z)\ge c^*>v(a_0)$, and thus $c^*\in\mathbb{R}$. Because $v$ is continuous and $X$ is connected, $v(X)$ is also a connected set in $\mathbb{R}$. In $\mathbb{R}$, connectedness is equivalent to convexity. Therefore, $v(X)$ is convex, and thus there exists $z^*\in X$ such that $v(z^*)=c^*$. By Lemma 4, there exists $w^*\in X$ such that $[w^*,a_0]=[z^*,w^*]$ and $z^*\succ w^*\succ a_0$. Because $c^*=v(z^*)>v(w^*)>v(a_0)$, there exists $k$ such that
\[v(a_{k+1})>v(w^*)\ge v(a_k).\]
Then, by (\ref{EQUIV}),
\[[a_{2k+2},a_{k+1}]=[a_{k+1},a_0]>[w^*,a_0]=[z^*,w^*]>[z^*,a_{k+1}],\]
and thus, by consistency,
\[a_{2k+2}\succ z^*.\]
This implies that $v(a_{2k+2})>c^*$, which contradicts the definition of $c^*$. This completes the proof. $\blacksquare$

\vspace{12pt}
Now, suppose that $x\sim y$ for every $x,y\in X$. Then, $[x,y]=[z,y]=[z,w]$ for every $x,y,z,w\in X$, and thus $u:X\to \mathbb{R}$ represents this system if and only if $u$ is a constant function. Hence, we hereafter assume that there exists $x^*,y^*\in X$ such that $x^*\succ y^*$.

We will recursively construct (possibly finite or infinite) sequences $(a_i^k)$ for $k\in \mathbb{N}$.\footnote{Later, we construct our utility function $u$ to satisfy $u(a_i^k)=\frac{i}{2^k}$.} First, we define $a_0^0=y^*$ and $a_1^0=x^*$. Suppose that $a_i^0$ is defined for some $i\ge 1$. If there exists $x\in X$ such that $[x,a_i^0]\ge [a_1^0,a_0^0]$, then by Lemma 2, there exists $a_{i+1}^0\in X$ such that $[a_{i+1}^0,a_i^0]=[a_1^0,a_0^0]$. If there is no such $x$, then we do not define $a_{i+1}^0$. Similarly, suppose that $a_i^0$ is defined for some $i\le 0$. If there exists $x\in X$ such that $[a_i^0,x]\ge [a_1^0,a_0^0]$, then by Lemma 3, there exists $a_{i-1}^0\in X$ such that $[a_i^0,a_{i-1}^0]=[a_1^0,a_0^0]$. If there is no such $x$, then we do not define $a_{i-1}^0$.

Next, suppose that $(a_i^k)$ is defined for $k=0,...,k^*$ and the following two relationships hold: 1) $a_i^k=a_{2i}^{k+1}$ and 2) $[a_{i+1}^k,a_i^k]=[a_1^k,a_0^k]$. We define a sequence $(a_i^{k^*+1})$ as follows. First, if $a_i^{k^*}$ is defined, then we define $a_{2i}^{k^*+1}=a_i^{k^*}$. Second, if both $a_i^{k^*}, a_{i+1}^{k^*}$ are defined, then by Lemma 4, there exists $a_{2i+1}^{k^*+1}\in X$ such that $[a_{2i+1}^{k^*+1},a_{2i}^{k^*+1}]=[a_{2i+2}^{k^*+1},a_{2i+1}^{k^*+1}]$. Third, suppose that $a_i^{k^*}$ is defined and $a_{i+1}^{k^*}$ is undefined. Note that, in this case we have $i\ge 2^{k^*}$. If there exists $x\in X$ such that $[x,a_i^{k^*}]\ge [a_1^{k^*+1},a_0^{k^*+1}]$, then by Lemma 2, there exists $a_{2i+1}^{k^*+1}\in X$ such that $[a_{2i+1}^{k^*+1},a_{2i}^{k^*+1}]=[a_1^{k^*+1},a_0^{k^*+1}]$. If such an $x$ does not exist, then we do not define $a_{2i+1}^{k^*+1}$.\footnote{Note that, if $a_{2i+1}^{k^*+1}$ is defined and $a_{i+1}^{k^*}$ is undefined, then there is no $x\in X$ such that $[x,a_{2i+1}^{k^*+1}]\ge [a_1^{k^*+1},a_0^{k^*+1}]$. Suppose that such an $x$ exists. Then, we can define $a_{2i+2}^{k^*+1}$. By the same arguments as in the proof of Lemma 5, we can show that (\ref{EQUIV}) holds for $(a_i^{k^*+1})$, and thus,
\[[a_1^{k^*},a_0^{k^*}]=[a_{2i+2}^{k^*+1},a_i^{k^*}],\]
which contradicts the undefinedness of $a_{i+1}^{k^*}$.} Similarly, suppose that $a_i^{k^*}$ is defined and $a_{i-1}^{k^*}$ is undefined. Then, we have that $i\le 0$. If there exists $x\in X$ such that $[a_i^{k^*},x]\ge [a_1^{k^*+1},a_0^{k^*+1}]$, then by Lemma 3, there exists $a_{2i-1}^{k^*+1}\in X$ such that $[a_{2i}^{k^*+1},a_{2i-1}^{k^*+1}]=[a_1^{k^*+1},a_0^{k^*+1}]$. If such an $x$ does not exist, then we do not define $a_{2i-1}^{k^*+1}$.

By the above arguments, we can define $(a_i^k)$ recursively, and these sequences satisfy $a_i^k=a_{2i}^{k+1}$ and $[a_{i+1}^k,a_i^k]=[a_1^k,a_0^k]$. Let $D\subset X$ be the set of all $a_i^k$. We need the following lemma.

\vspace{12pt}
\noindent
\textbf{Lemma 6}. If $x\succ y$, then there exists $z\in D$ such that $x\succ z\succ y$.\footnote{If there is a representation $u:X\to \mathbb{R}$ such that $u(x^*)=1$ and $u(y^*)=0$, then this statement says that there exists $k\ge 1$ and $i\in \mathbb{Z}$ such that $u(x)>\frac{i}{2^k}>u(y)$.}

\vspace{12pt}
\noindent
\textbf{Proof}. Suppose that $x\succ y$ and there is no $z\in D$ such that $x\succ z\succ y$. We treat the case in which $x\succ a_0^0$ because the remaining case can be treated symmetrically.

Because of our assumption, we must have $y\succsim a_0^0$. First, we assume that there exists $z\in D$ such that $y\succsim z\succ a_0^0$. By Lemma 5, there exists a finite sequence $b_1,b_0,...,b_{-k}$ such that $b_1=x$, $b_0=y$, $[b_{i+1},b_i]=[b_1,b_0]$ for all $i\in \{-k,...,-1\}$, $b_{-k}\succsim a_0^0$ and $[b_1,b_0]>[b_{-k},a_0^0]$. On the other hand, again by Lemma 5, for every $m\ge 0$, there exists $\ell\ge 0$ such that $y\succsim a_{\ell}^m$ and $[a_1^m,a_0^m]>[y,a_{\ell}^m]$. If there does not exist $a_{\ell+1}^m$, then by our definition of $(a_i^m)$, $[a_1^m,a_0^m]>[x,a_{\ell}^m]$. If $a_{\ell+1}^m$ exists, then $a_{\ell+1}^m\succ y$. Because $a_{\ell+1}^m\in D$ and there is no $z\in D$ such that $x\succ z\succ y$, we have that $a_{\ell+1}^m\succsim x$, and thus
\[[a_1^m,a_0^m]=[a_{\ell+1}^m,a_{\ell}^m]\ge [x,a_{\ell}^m].\]
Therefore, in any case, we have that $[a_1^m,a_0^m]\ge [x,a_{\ell}^m]$.

Because there exists $z\in D$ such that $y\succsim z\succ a_0^0$, $\ell\ge 1$ for some $m$. Because $a_{\ell}^m=a_{2\ell}^{m+1}$, we can choose $m$ so large that $\ell>k$. Define $c_0=a_{\ell}^m$. Then, clearly $b_0\succsim c_0\succsim a_{\ell}^m$. Note that,
\[[a_1^m,a_0^m]\ge [x,a_{\ell}^m]\ge [x,y]=[b_1,b_0].\]
On the other hand,
\[[b_0,a_{\ell-1}^m]\ge [a_{\ell}^m,a_{\ell-1}^m]=[a_1^m,a_0^m]\ge [b_0,b_0],\]
and thus, by Lemma 3, there exists $c_1\in X$ such that
\[[b_0,c_1]=[a_1^m,a_0^m].\]
If $k\ge 1$, then 
\[[b_0,c_1]\ge [b_1,b_0]=[b_0,b_{-1}],\]
and thus, $b_{-1}\succsim c_1\succsim a_{\ell-1}^m$. Inductively, we can show that there exists $c_k\in X$ such that $b_{-k}\succsim c_k\succsim a_{\ell-k}^m$. Then,
\[[b_{-k},a_{\ell-k-1}^m]\ge [a_{\ell-k}^m,a_{\ell-k-1}^m]\ge [x,y]\ge [b_{-k},b_{-k}],\]
and thus, there exists $b_{-k-1}\in X$ such that $b_{-k-1}\succsim a_{\ell-k-1}^m\succsim a_0^0$ and
\[[b_{-k},b_{-k-1}]=[x,y],\]
which contradicts our definition of $k$.

Therefore, we can assume that there is no $z\in D$ such that $y\succsim z\succ a_0^0$, and thus, there is no $z\in D$ such that $x\succ z\succ a_0^0$. By Lemma 5, there exists a finite sequence $b_0,...,b_k$ such that
\[b_0=a_0^0,\ b_1=x,\ [b_{i+1},b_i]=[b_{i+2},b_{i+1}],\ a_1^0\succsim b_k,\ [b_1,b_0]>[a_1^0,b_k].\]
Choose any $m$ such that $k+1<2^m$. Then, $a_1^0=a_{2^m}^m$. Suppose that $a_{2^m-1}^m\succsim b_k$. Then,
\[[b_1,b_0]>[a_{2^m}^m,b_k]\ge [a_{2^m}^m,a_{2^m-1}^m]=[a_1^m,a_0^m],\]
and thus, $b_1\succ a_1^m\succ b_0$, which contradicts our initial assumption. Therefore, $b_k\succ a_{2^m-1}^m$. Because $k+1<2^m$, by the pigeonhole principle, there exist $i,j$ such that $b_{i+1}\succ a_{j+1}^m\succ a_j^m\succsim b_i$. Then,
\[[b_1,b_0]=[b_{i+1},b_i]>[a_{j+1}^m,b_i]\ge[a_{j+1}^m,a_j^m]=[a_1^m,a_0^m],\]
and thus $b_1\succ a_1^m\succ b_0$, which is a contradiction. This completes the proof. $\blacksquare$

\vspace{12pt}
Define
\[D'=\{x\in X|\exists x'\in D\mbox{ s.t. }x\sim x'\}.\]
Let $x\in D'$. Then, there exist $i,k$ such that $x\sim a_i^k$. Define
\[u(x)=\frac{i}{2^k}.\]
Clearly, $u(x)$ does not depend on the choice of $k$. Choose any $x,y,z,w\in D'$. Then, there exist $k$ and $i_x,i_y,i_z,i_w$ such that
\[x\sim a_{i_x}^k,\ y\sim a_{i_y}^k,\ z\sim a_{i_z}^k,\ w\sim a_{i_w}^k.\]
Suppose that
\[u(x)-u(y)\ge u(z)-u(w).\]
This means that
\[i_x-i_y\ge i_z-i_w.\]
If $i_x-i_y\ge 0\ge i_z-i_w$, then
\[[x,y]\ge [y,y]=[z,z]\ge [z,w].\]
If $i_x-i_y\ge i_z-i_w\ge 0$, define $i_*=\min\{i_y,i_w\}$. Then, both $a_{i_*+i_x-i_y}^k$ and $a_{i_*+i_z-i_w}^k$ are defined, and by (\ref{EQUIV}),\footnote{Note that, we can show that (\ref{EQUIV}) holds for $(a_i^m)$ by the same arguments as in the proof of Lemma 5.}
\[[x,y]=[a_{i_*+i_x-i_y},a_{i_*}^k]\ge [a_{i_*+i_z-i_w},a_{i_*}^k]=[z,w].\]
By the symmetrical arguments, we can show that if $0\ge i_x-i_y\ge i_z-i_w$, then $[x,y]\ge [z,w]$. In conclusion, $u(x)-u(y)\ge u(z)-u(w)$ implies $[x,y]\ge [z,w]$. Moreover, if 
\[u(x)-u(y)>u(z)-u(w),\]
then $i_x-i_y>i_z-i_w$, and by repeating the above arguments, we have that
\[[x,y]>[z,w].\]
Therefore, $u$ represents $\ge$ on $D'$.

We extend the definition of $u$ on $X$. If $x$ is not a least element of $X$ on $\succsim$, then by Lemma 6, there exists $y\in D$ such that $x\succ y$. Define
\[u(x)=\sup\{u(y)|y\in D,\ x\succsim y\}.\]
Conversely, if $x$ is not a greatest element of $X$ on $\succsim$, then by Lemma 6, there exists $y\in D$ such that $y\succ x$. Define
\[u(x)=\inf\{u(y)|y\in D,\ y\succsim x\}.\]
We can easily show that these two definitions of $u$ coincide at each $x$ such that both can be defined.

If $x\succsim y$, then clearly $u(x)\ge u(y)$. If $x\succ y$, then by Lemma 6, there exists $z,w\in D$ such that $x\succ z\succ w\succ y$. Therefore,
\[u(x)\ge u(z)>u(w)\ge u(y),\]
and thus $u(x)>u(y)$. Hence,
\[x\succsim y\Leftrightarrow u(x)\ge u(y).\]

Next, we show the continuity of $u$. For this purpose, we should show that both $u^{-1}(]-\infty,a])$ and $u^{-1}([a,+\infty[)$ are closed for every $a\in \mathbb{R}$. If $\inf u<a<\sup u$, then
\[u^{-1}([a,+\infty[)=X\setminus \cup_{y\in D: u(y)<a}\{z\in X|y\succ z\},\]
\[u^{-1}(]-\infty,a])=X\setminus \cup_{y\in D: u(y)>a}\{z\in X|z\succ y\},\]
and both are clearly closed. The case in which either $a\le \inf u$ or $\sup u\le a$ can be treated easily, and thus we omit the proof for such cases.

To prove that $u$ represents $\ge$, it suffices to show that
\[[x,y]\ge [z,w]\Leftrightarrow u(x)-u(y)\ge u(z)-u(w).\]
For this, it suffices to show the following two claims:
\begin{enumerate}[i)]
\item if $u(x)-u(y)\ge u(z)-u(w)$, then $[x,y]\ge [z,w]$, and

\item if $u(x)-u(y)>u(z)-u(w)$, then $[x,y]>[z,w]$.
\end{enumerate}
We show that i) implies ii). Suppose that i) is correct. Choose any $x,y,z,w\in X$ such that $u(x)-u(y)>u(z)-u(w)$. Note that, because $X$ is connected, $u(X)$ is also connected in $\mathbb{R}$, and thus it is convex. If $u(x)>\inf u$, then there exists $x'\in X$ such that $u(x')-u(y)\ge u(z)-u(w)$ and $x\succ x'$. Then,
\[[x,y]>[x',y]\ge [z,w],\]
which implies that $[x,y]>[z,w]$. If $u(x)=\inf u$, then $w>\inf u$ because $u(x)+u(w)>u(y)+u(z)$. Therefore, there exists $w'$ such that $u(x)-u(y)\ge u(z)-u(w')$ and $w\succ w'$. Then,
\[[x,y]\ge [z,w']>[z,w],\]
which implies that $[x,y]>[z,w]$. Therefore, our claim is correct, and thus it suffices to show i).

We need the following lemma.

\vspace{12pt}
\noindent
\textbf{Lemma 7}. Suppose that $x\in X$. Then, there exist $x'\in X$ and a sequence $(x^m)$ on $D'$ such that $u(x)=u(x')$ and $x^m\to x'$ as $m\to \infty$. Moreover, if $u(x)<\sup u$ (resp. $u(x)>\inf u$), then we can assume that $u(x^m)>u(x)$ (resp. $u(x^m)<u(x)$) for every $m$.\footnote{In the proof of Lemma 7, we need the path-connectedness of $X$. In proving Theorem 1, only Lemma 7 requires the path-connectedness of $X$, and the remaining part uses only connectedness. Therefore, if there is a proof of Lemma 7 that does not use the path-connectedness of $X$, we can extend Theorem 1 to a general Hausdorff, separable, and connected topological space.}

\vspace{12pt}
\noindent
\textbf{Proof}. We treat only the case in which $u(x)<\sup u$, because the remaining case can be treated symmetrically.

By assumption, there exists $\hat{x}\in X$ such that $u(\hat{x})>u(x)$. Because $X$ is path-connected, there exists a continuous function $f:[0,1]\to X$ such that $f(0)=x, f(1)=\hat{x}$. Define $t'=\sup\{t\in [0,1]|u(f(t))\le u(x)\}$ and $x'=f(t')$. Clearly, $t'<1$ and $u(x')=u(x)$. Let $g(t)=u(f(t))$. Then, we have that if $t'<t<1$, then $g(t)>g(t')=v(x)$. Therefore, by the intermediate value theorem, there exists $t_1$ such that $t'<t_1<1$ and $f(t_1)\in D'$. Define $x^1=f(t_1)$. Next, suppose that $t_m$ and $x_m=f(t_m)$ are defined, where $t_m>t'$ and $x_m\in D'$. Because $t_m>t'$, $t'<t<t_m$ implies $g(t)>u(x)$. Therefore, again by the intermediate value theorem, there exists $t_{m+1}$ such that $t'<t_{m+1}<\frac{t'+t_m}{2}$ and $x_{m+1}=f(t_{m+1})\in D'$. By construction,
\[0<t_m-t'<(t_1-t')/2^{m-1},\]
and thus $t_m\to t'$ as $m\to \infty$. By the continuity of $f$, $x^m\to x'$ as $m\to \infty$. This completes the proof. $\blacksquare$

\vspace{12pt}
Assume that
\[u(x)-u(y)\ge u(z)-u(w).\]
Then,
\[u(x)+u(w)\ge u(y)+u(z)\]
If $u(y)=u(z)=\inf u$, then clearly
\[[x,y]\ge [x,w]\ge [z,w].\]
If $u(x)=u(w)=\sup u$, then clearly
\[[x,y]\ge [z,y]\ge [z,w].\]
Therefore, without loss of generality, we can assume that either $u(x)<\sup u$ or $u(w)<\sup u$, and either $u(y)>\inf u$ or $u(z)>\inf u$. We treat the case in which $u(x)<\sup u$ and $u(y)>\inf u$, because the remaining cases can be treated similarly.

Because $u(x)<\sup u$ and $u(y)>\inf u$, by Lemma 7, there exist $x',y',z',w'\in X$ and sequences $(x^m), (y^m), (z^m), (w^m)$ on $D'$ such that $u(x^m)>u(x)$ and $u(y^m)<u(y)$ for all $m$,
\[u(x')=u(x),\ u(y')=u(y),\ u(z')=u(z),\ u(w')=u(w),\]
and
\[x^m\to x',\ y^m\to y',\ z^m\to z',\ w^m\to w'\mbox{ as }m\to \infty.\]
Then, for every $m$,
\[u(x^m)-u(y^m)>u(z)-u(w),\]
and thus, there exists $k(m)\ge m$ such that
\[u(x^m)-u(y^m)\ge u(z^{k(m)})-u(w^{k(m)}).\]
Because $u$ represents $\ge$ on $D'$,
\[[x^m,y^m]\ge [z^{k(m)},w^{k(m)}].\]
Therefore, by continuity,
\[[x,y]=[x',y']\ge [z',w']=[z,w].\]
This implies i). Hence, $u$ represents $\ge$ on $X$.

The rest of our claim is the uniqueness of such a $u$ up to a positive affine transform. Suppose that $v:X\to \mathbb{R}$ is a continuous representation of $\ge$. If $x\sim y$ for all $x,y\in X$, then $v$ is a constant function, in which case the uniqueness is obvious. Hence, we assume that there exist $x^*,y^*\in X$ such that $x^*\succ y^*$. Define $u:X\to \mathbb{R}$ as in the above proof. Then, $u$ is a representation of $\ge$ such that $u(a_i^k)=\frac{i}{2^k}$ for all $a_i^k\in D$. We can easily show that
\begin{align*}
v(a_i^k)=&~v(a_i^k)-v(a_0^k)+v(y^*)\\
=&~i(v(a_1^k)-v(a_0^k))+v(y^*)\\
=&~\frac{i}{2^k}(v(x^*)-v(y^*))+v(y^*)\\
=&~(v(x^*)-v(y^*))u(a_i^k)+v(y^*).
\end{align*}
Because $v$ is continuous, by Lemma 7,
\[v(x)=(v(x^*)-v(y^*))u(x)+v(y^*),\]
as desired. This completes the proof. $\blacksquare$

\subsection{Proof of Theorem 3}
Suppose that $u$ is continuously differentiable on $\mathbb{R}^n_{++}$ and $Du(x)\neq 0$ for every $x\in \mathbb{R}^n_{++}$. Define $v(x,y)=u(x)-u(y)$. Then,
\[I=\{(x,y)\in \mathbb{R}^{2n}_{++}|x\sim y\}=v^{-1}(0).\]
By the preimage theorem, $I$ is a $2n-1$ dimensional $C^1$ manifold, and thus Debreu's smoothness holds.\footnote{See section 1.4 of Guillemin and Pollack (1974).} Next, recall that $e=(1,1,...,1)$. Define $g(b)=u(be)$. Then, $g$ is a concave, increasing, and continuously differentiable function defined on $\mathbb{R}_{++}$, and because $Du(x)\neq 0$, $g'(b)=Du(be)e>0$. Therefore, by the inverse function theorem, $g^{-1}$ is also continuously differentiable. This implies that
\[\lim_{a\downarrow 0}\frac{b-f(a,b)}{a}=0\Leftrightarrow \lim_{a\downarrow 0}\frac{g(b)-g(f(a,b))}{a}=0.\]
Because $g$ is concave, $2g(b)\ge g(b+a)+g(b-a)$, and thus $b\ge f(a,b)\ge b-a$. This implies that
\[\liminf_{a\downarrow 0}\frac{g(b)-g(f(a,b))}{a}\ge 0.\]
On the other hand,
\begin{align*}
0\le&~\limsup_{a\downarrow 0}\frac{g(b)-g(f(a,b))}{a}\\
=&~\lim_{a\downarrow 0}\frac{g(b)-g(b-a)}{a}-\liminf_{a\downarrow 0}\frac{g(f(a,b))-g(b-a)}{a}\\
=&~\lim_{a\downarrow 0}\frac{g(b)-g(b-a)}{a}-\liminf_{a\downarrow 0}\frac{g(b+a)-g(f(a,b))}{a}\\
\le&~\lim_{a\downarrow 0}\frac{g(b)-g(b-a)}{a}-\liminf_{a\downarrow 0}\frac{g(b+a)-g(b)}{a}\\
=&~g'(b)-g'(b)=0,
\end{align*}
which implies that line smoothness is satisfied.

Conversely, suppose that $\ge$ satisfies both Debreu's smoothness and line smoothness. First, we show that the function
\[g(b)=u(be)\]
is continuously differentiable on $\mathbb{R}_{++}$. Choose any $b>0$. Because $g$ is concave, it is locally Lipschitz, and thus there exist $\varepsilon>0$ and $L>0$ such that $|g(c)-g(b)|\le L|c-b|$ for all $c\in [b-\varepsilon,b+\varepsilon]$. Moreover, the left- and right-side derivatives $D_-g(b), D_+g(b)$ can be defined and are real numbers, and $g$ is differentiable if and only if $D_-g(b)=D_+g(b)$. Because $2g(b)\ge g(b+a)+g(b-a)$, we have that $b-a\le f(b,a)\le b$ for all $a>0$, and thus by monotonicity, $g(b)\ge g(f(a,b))$. Moreover,
\begin{align*}
0\le&~\liminf_{a\downarrow 0}\frac{g(b)-g(f(a,b))}{a}\\
\le&~\limsup_{a\downarrow 0}\frac{g(b)-g(f(a,b))}{a}\\
\le&~L\lim_{a\downarrow 0}\frac{b-f(a,b)}{a}=0,
\end{align*}
and thus
\[\lim_{a\downarrow 0}\frac{g(b)-g(f(a,b))}{a}=0.\]
Therefore,
\begin{align*}
D_+g(b)=&~\lim_{a\downarrow 0}\frac{g(b+a)-g(b)}{a}\\
=&~\lim_{a\downarrow 0}\frac{g(b+a)-g(f(a,b))}{a}\\
=&~\lim_{a\downarrow 0}\frac{g(f(a,b))-g(b-a)}{a}\\
=&~\lim_{a\downarrow 0}\frac{g(b)-g(b-a)}{a}\\
=&~D_-g(b),
\end{align*}
as desired. Because every differentiable and concave function is continuously differentiable, $g$ is continuously differentiable.

Because $g$ is increasing and concave, we must have $g'(b)>0$ for every $b>0$.

Now, because $u$ is continuous and increasing, for every $x\in \mathbb{R}^n_{++}$, there uniquely exists $a(x)>0$ such that $u(x)=u(a(x)e)=g(a(x))$. Therefore, it suffices to show that, under monotonicity and Debreu's smoothness, $a:x\mapsto a(x)$ is continuously differentiable and $Da(x)\neq 0$ on $\mathbb{R}^n_{++}$. We treat this result as a lemma.\footnote{If $I$ is a $C^2$ manifold, then this result was obtained by Debreu (1972), although his proof has a gap. If $I$ is a $C^{\infty}$ manifold, then this result was rigorously proved in ch.7 of Bridges and Mehta (1995). However, we can only assume that $I$ is $C^1$, and thus these results cannot be applied. Mas-Colell (1977) noted that Moulin solved this problem in 1973. However, we could not obtain Moulin's monograph, and thus this claim is unverifiable. This is why we present the proof of the smoothness of $a(x)$.}

\vspace{12pt}
\noindent
\textbf{Lemma 8}. Suppose that $X$ is either $\mathbb{R}^n_+$ or $\mathbb{R}^n_{++}$, and $\succsim$ is a closed and monotone weak order on $X$,\footnote{We say that a weak order $\succsim$ on $X$ is {\bf monotone} if $x\gg y$ implies that $x\succ y$.} and the set
\[I=\{(x,y)\in\mathbb{R}^{2n}_{++}|x\sim y\}\]
is $2n-1$ dimensional $C^k$ manifold. Then, for every $x\in \mathbb{R}^n_{++}$, there uniquely exists $a(x)$ such that $x\sim a(x)e$, where $e=(1,1,...,1)$. Moreover, the function $a:\mathbb{R}^n_{++}\to\mathbb{R}_{++}$ is $C^k$, and $Da(x)\neq 0$ for every $x\in \mathbb{R}^n_{++}$.

\vspace{12pt}
\noindent
\textbf{Proof}. For $x\in \mathbb{R}^n_{++}$, define $a(x)=\inf\{c>0|ce\succsim x\}$. It is easy to show that $a(x)$ is the unique number $c>0$ such that $x\sim ce$, and thus we omit its proof.\footnote{See Proposition 3.C.1 of Mas-Colell et al. (1995).}

Choose any $x\in\mathbb{R}^n_{++}$, and define
\[I_x=\{y\in \mathbb{R}^n_{++}|x\sim y\}.\]
We first show that $I_x$ is connected. For this, it suffices to show that, for every $y\in I_x$, there exists a continuous function $f:[0,2]\to I_x$ such that $f(0)=x$ and $f(2)=y$.

Define $z_i=\min\{x_i,y_i\}>0$, and 
\[b(t)=\begin{cases}
(1-t)x+tz & \mbox{if }0\le t\le 1,\\
(2-t)z+(1+t)y & \mbox{if }1\le t\le 2.
\end{cases}\]
Let $M=\max_i\{|x_i-y_i|\}$. By monotonicity and the intermediate value theorem, there uniquely exists $c(t)\in [0,M]$ such that $b(t)+c(t)e\in I_x$. Let $(t_m)$ be the sequence in $[0,2]$ such that $t_m\to t$ as $m\to \infty$. Suppose that $c(t_m)\not\to c(t)$. Because $c(t_m)\in [0,M]$, there exists a subsequence $t_{\ell(m)}$ such that $c(t_{\ell(m)})\to \alpha\neq c(t)$ as $m\to \infty$. Because $\succsim$ is closed in $X^2$, $b(t)+\alpha e\in I_x$, which contradicts the uniqueness of $c(t)$. Therefore, $c(t)$ is continuous, and $f(t)=b(t)+c(t)e$ satisfies all of our requirements.

For $x\in \mathbb{R}^n_{++}$, define $L_x$ as the set of all $y\in I_x$ such that the function $a:\mathbb{R}^n_{++}\to \mathbb{R}_{++}$ is $C^k$ around $y$ and $Da(y)\neq 0$. To prove this lemma, it suffices to show that $x\in L_x$ for all $x\in \mathbb{R}^n_{++}$. Fix $x\in \mathbb{R}^n_{++}$. To show that $x\in L_x$, it suffices to show that $L_x=I_x$. Because $I_x$ is connected, it suffices to show that $L_x$ is nonempty, open, and closed with respect to the relative topology of $I_x$.

It is clear that $L_x$ is open.

Choose any $y\in I_x$. Then, $a(y)=a(x)$. Because $I$ is a $2n-1$ dimensional $C^k$ manifold, there exist open neighborhoods $U\subset \mathbb{R}^{2n-1}$ of $0$ and $V\subset I$ of $(y,y)$, and a $C^k$ diffeomorphism $\varphi:U\to V$ such that $\varphi(0)=(y,y)$.\footnote{The statement ``$\varphi$ is a $C^k$ diffeomorphism'' means that $\varphi$ is a bijection and both $\varphi$ and $\varphi^{-1}$ are $C^k$. Note that, $V$ is open not in $\mathbb{R}^{2n}$ but only in $I$, and thus to understand this statement rigorously, we must define the differentiability of functions defined on non-open sets. See section 1.1-1.2 of Guillemin and Pollack (1974).} Let $T_{(y,y)}(I)$ be the image of $\mathbb{R}^{2n-1}$ for the linear operator $D\varphi(0)$. This linear space is called the \textbf{tangent space} of $I$ at $(y,y)$. It is known that $v\in T_{(y,y)}(I)$ if and only if there exists a continuously differentiable function $c:[-\varepsilon,\varepsilon]\to I$ such that $c(0)=(y,y)$ and $\dot{c}(0)=v$. Because $c(t)=(y+te,y+te)\in I$, we have that $(e,e)\in T_{(y,y)}(I)$. If $(0,e)\in T_{(y,y)}(I)$, then 
\[(-e,e)=2(0,e)-(e,e)\in T_{(y,y)}(I),\]
and thus, there exists a continuously differentiable function $c:[-\varepsilon,\varepsilon]\to I$ such that $\dot{c}(0)=(-e,e)$. This means that there exists $(z,w)\in I$ such that $w\gg y$ and $y\gg z$. However, by monotonicity,
\[w\succ y\succ z\sim w,\]
which is a contradiction. Define
\[\psi(v,a)=\varphi(v)+(0,ae).\]
Then, by the above result, $D\psi(0,0)$ is a regular matrix, and thus, by the inverse function theorem, there exist open neighborhoods $U'\subset \mathbb{R}^{2n}$ of $(0,0)$ and $V'\subset \mathbb{R}^{2n}_{++}$ of $(y,y)$ such that $\psi:U'\to V'$ is a $C^k$ diffeomorphism. Note that for $(z,w)\in V'$, $z\sim w$ if and only if $\psi^{-1}(z,w)=(v,0)$ for some $v$. Let $\chi(z,w)=\psi^{-1}(z,w)$, and $h(z,b)=\chi_{2n}(z,y+be)$. Then, $h$ is $C^k$ and $z\sim (y+be)$ if and only if $h(z,b)=0$. 

We show that $D_yh(y,0)\neq 0$ and $\frac{\partial h}{\partial b}(y,0)\neq 0$. Because $h(y+be,b)=0$ for all $b$, $D_yh(y,0)e=-\frac{\partial h}{\partial b}(y,0)$, and thus, it suffices to show that $\frac{\partial h}{\partial b}(y,0)\neq 0$. Suppose that $\frac{\partial h}{\partial b}(y,0)=0$. This implies that
\[D\chi_{2n}(y,y)(0,e)=0.\]
Meanwhile,
\[(0,e)=D\psi(0,0)D\chi(y,y)(0,e),\]
and
\[D\psi(0,0)(w,b)=D\varphi(0)w+b(0,e).\]
Therefore,
\[(0,e)=D\varphi(0)w\]
for some $w\in \mathbb{R}^{2n-1}$. This implies that $(0,e)\in T_{(y,y)}(I)$, which contradicts our previous arguments. Therefore, our claim is correct. By the implicit function theorem, there exists an open neighborhood $W'$ of $y$ and a $C^k$ function $b:W'\to \mathbb{R}$ such that $z\sim y+b(z)e$ and $Db(z)\neq 0$ for every $z\in W'$, and $b(y)=0$.

If $y=a(x)e$, then $b(z)$ coincides with $a(z)-a(x)$, and thus $a(x)e\in L_x$ and $L_x$ is nonempty.

Thus, it suffices to show that $L_x$ is closed in $I_x$. Choose a sequence $(y^m)$ on $L_x$ and suppose that $y^m\to y\in I_x$ as $m\to \infty$. Consider the function $b:W'\to \mathbb{R}$ defined as above. Because of monotonicity, $b$ is increasing, and thus $Db(z)e>0$ for every $z\in W'$. Because $y^m\to y$ as $m\to \infty$, there exists $m$ such that $y^m\in W'$. Consider the following equation:
\[b(z)-b(y^m+te)=0.\]
This equation holds for $(z,t)=(y,0)$. By the implicit function theorem, there exist an open neighborhood $W''\subset W'$ of $y$ and a $C^k$ function $t:W''\to \mathbb{R}$ such that $Dt(z)e\neq 0$ and $b(z)=b(y^m+t(z)e)$ for every $z\in W''$, and $t(y)=0$. This implies that
\[a(z)=a(y^m+t(z)e).\]
Because $y^m\in L_x$, $a$ is $C^k$ around $y$ and $Da(y)\neq 0$, which implies that $y\in L_x$ and $L_x$ is closed. This completes the proof. $\blacksquare$

\vspace{12pt}
Because of Lemma 8, $u$ satisfies all of our requirements. This completes the proof. $\blacksquare$

\section*{Acknowledgement}
The author would like to express the gratitude to Marcel K. Richter. In 2013, the author met Marcel at Keio University, and was presented with some important pieces of advice and suggestions. All of these have been immensely worthwhile. Unfortunately, Marcel died in 2014. The author prays for the repose of his soul. The author would also like to thank Mitsunobu Miyake for many suggestions at the 2015 DC Conference and the 2020 Japan Economic Association Fall Meeting. The author is also grateful to Chaowen Yu, Shinsuke Nakamura, and Shinichi Suda for their helpful comments and suggestions. This work was supported by JSPS KAKENHI Grant Number JP21K01403. Finally, the author thanks the anonymous reviewers for their helpful comments and suggestions on this paper.

\section*{References}

\begin{description}
\item{[1]} Alexandrov, A. D. 1939. Almost Everywhere Existence of the Second Differential of a Convex Function and Some Properties of Convex Surfaces Connected with It. Leningrad State Univ. Ann. Math. Ser. 6, 3-35.

\item{[2]} Alt, F. 1936. \"Uber die Messbarkeit des Nutzens. Zeitschrift f\"ur National\"okonomie 7, 161-169.

\item{[3]} von Auspitz, R., Lieben, R. 1889. Untersuchungen \"uber die Theorie des Preises. Verlag von Duncker und Humboldt, Leibzig.

\item{[4]} Bridges, D. S., Mehta, G. B. 1995. Representations of Preference Orderings. Springer, Berlin.

\item{[5]} Chipman, J. S., Hurwicz, L., Richter, M. K., Sonnenschein, H. F. 1971. Preferences, Utility and Demand. Harcourt Brace Jovanovich, Inc., New York.

\item{[6]} Debreu, G. 1954. Representation of a Preference Ordering by a Numerical Function.'' In: Thrall, R. M., Coombs, C. H., Davis R. L. (Eds.) Decision Processes. Wiley, pp.159-165.

\item{[7]} Debreu, G. 1972. Smooth Preferences. Econometrica 40, 603-615.

\item{[8]} Debreu, G. 1976. Least Concave Utility Function. J Math. Econ. 3, 121-129.

\item{[9]} Edgeworth, F. Y. 1897. La Teoria Pura del Monopolio. Giornale degli Economisti Serie Seconda 15, 13-31.

\item{[10]} Epstein, L. G., Zhang, J. 1999. Least Convex Capacities. Econ. Theory 13, 263-286.

\item{[11]} Gerasimou, G. 2021. Simple Preference Intensity Comparisons. Forthcoming to J Econ. Theory 192.

\item{[12]} Guillemin, V., Pollack, A. 1974. Differential Topology. Prentice Hall, New Jersey.

\item{[13]} Howard, R. 1998. Alexandrov's Theorem on the Second Derivatives of Convex Functions via Rademacher's Theorem on the First Derivatives of Lipschitz Functions. Lecture note from a functional analysis seminar at the University of South Carolina.

\item{[14]} Ioffe, A. D., Tikhomirov, V. M. 1979. Theory of Extremal Problem. North Holland, Amsterdam.

\item{[15]} Kannai, Y. 1977. Concavifiability and Constructions of Concave Utility Functions. J Math. Econ. 4, 1-56.

\item{[16]} Kannai, Y. 1980. The ALEP Definition of Complementarity and Least Concave Utility Function. J Econ. Theory 22, 115-117.

\item{[17]} Kawamata, K. 2010. Obituary: Professor Paul A. Samuelson. Keio J Econ. 103, 299-307.

\item{[18]} Kranz D. H., Luce, R. D., Suppes, P., Tversky, A. 1971. Foundations of Measurement, Vol. 1, Academic Press, New York. 

\item{[19]} Luce, R. D., Raiffa, H. 1957. Games and Decisions: Introduction and Critical Survey. Wiley, New York.

\item{[20]} Mas-Colell, A. 1977. The Recoverability of Consumers' Preferences from Market Demand Behavior. Econometrica 45, 1409-1430.

\item{[21]} Mas-Colell, A., Whinston, M. D., Green, J. R. 1995. Microeconomic Theory. Oxford University Press, Oxford.

\item{[22]} Miyake, M. 2016. Logarithmically Homogeneous Preferences. J Math. Econ. 67, 1-9.

\item{[23]} Niculescu, C. P., Persson, L-E. 2018, Convex Functions and Their Applications. Springer International Publishing AG, Cham.

\item{[24]} Pareto, V. 1906. Manuale di Economia Politica con una Introduzione alla Scienza Sociale. Societa Editrice Libraria, Milano.

\item{[25]} Rockafeller, R. T. 1996. Convex Analysis. Princeton University Press, Princeton.

\item{[26]} Samuelson, P. A. 1974. Complementarity: An Essay on The 40th Anniversary of the Hicks-Allen Revolution in Demand Theory. J Econ. Lit. 12, 1255-1289.

\item{[27]} Seidl, C., Schmidt, U. 1997. Pareto on Intra- and Interpersonal Comparability of Utility. Hist. Econ. Ideas 5, 19-33.

\item{[28]} Shapley, L. S. 1975. Cardinal Utility from Intensity Comparisons. Report R-1683-PR, The Rand Corporation, Santa Monica, CA.

\item{[29]} Wakker, P. 1988. The Algebraic versus the Topological Approach to Additive Representations. J Math. Psychol. 32, 421-435.

\end{description}

\end{document}